\documentclass[journal]{IEEEtran}

\usepackage{cite}
\usepackage{array}
\usepackage{color}
\usepackage{float}
\usepackage{stfloats}
\usepackage[cmex10]{amsmath}
\usepackage{amsthm}
\usepackage{nomencl}						
\usepackage[normalem]{ulem}			        
\usepackage{fixltx2e} 					    
\usepackage{diagbox}						
\usepackage{slashbox}						
\usepackage{colortbl}						
\usepackage{multirow}						
\usepackage{tabularx}
\usepackage{placeins}
\usepackage{algorithm}
\usepackage{mathrsfs}
\usepackage{mathdots}
\usepackage{amssymb}
\usepackage{arydshln}
\usepackage{color}
\usepackage{gensymb}
\usepackage[noend]{algpseudocode}
\usepackage{bm}
\usepackage{url}
\usepackage[hidelinks]{hyperref}
\usepackage[T1]{fontenc}
\usepackage{cuted}
\usepackage{geometry}
\usepackage{amsmath}

\ifCLASSINFOpdf
  \usepackage[pdftex]{graphicx}
	\graphicspath{{./figure/}}
  \DeclareGraphicsExtensions{.pdf,.jpeg,.png}
\else
  \usepackage[dvips]{graphicx}
  \graphicspath{{./figure/}}
  \DeclareGraphicsExtensions{.eps}
\fi

\ifCLASSOPTIONcompsoc
  \usepackage[caption=false,font=normalsize,labelfont=sf,textfont=sf]{subfig}
\else
  \usepackage[caption=false,font=footnotesize]{subfig}
\fi

\ifCLASSOPTIONonecolumn

\else

\fi

\newtheorem{theorem}{Theorem}
\newtheorem{remark}{Remark}

\newcommand{\figref}[1]{\textcolor{black}{Fig.~\ref{#1}}}

\newcommand{\sectionref}[1]{\textcolor{black}{Section~\ref{#1}}}

\begin{document}
\bstctlcite{IEEEexample:BSTcontrol}
\setcounter{page}{1}

\title{Whole-System First-Swing Stability of Inverter-Based Inertia-Free Power Systems}

\author{Yitong~Li, \IEEEmembership{Member, IEEE}, Yunjie~Gu, \IEEEmembership{Senoir Member, IEEE}}

\ifCLASSOPTIONpeerreview
	\maketitle 
\else
	\maketitle
\fi

\begin{abstract}
The emphasis on inertia for system stability has been a long-held tradition in conventional grids. The fast and flexible controllability of inverters opens up new possibilities. This paper investigates the {first-swing stability} of inverter-based inertia-free power systems. We illustrate that by replacing inertia with fast primary control (i.e. inertia-free), the first-swing stability region is greatly extended compared to the classic equal area criterion of inertia-rich systems. The extended stability region is fully decentralised and ensures \textit{whole-system stability} for all swing modes across all timescales and interaction boundaries in a complex grid. The removal of inertia has impacts on frequency stability but these impacts can be well mitigated. The findings of the paper are proved mathematically and verified by simulation on the IEEE 68-bus system. 
\end{abstract}



\section{Introduction}

Rotating inertia has been the cornerstone of power system stability ever since the early days of ac electricity more than a hundred years ago \cite{kundur1994power,hatziargyriou2020definition}. The reasons behind this are twofold: (a) conventional generators have large intrinsic inertias due to the mechanical property of turbines; (b) the speed governors for turbines respond rather slowly and need inertias to serve as short-term energy buffers. When power systems are moving towards inverter-based renewable resources, none of the above two conditions holds. First, inverters do not have intrinsic inertias that are directly coupled to the grid \cite{markovic2021understanding,gu2022power,rocabert2012control,o2021enabling}. The renewable resources (e.g. wind turbines, photovoltaics, and batteries) are interfaced to the grid indirectly by inverters so their frequency regulation services can be shaped in a more flexible way, not necessarily limited to emulated inertias \cite{rocabert2012control,fang2018inertia,li2021revisiting}. Second, inverters respond much faster than speed governors, enabling much faster primary (droop) control, so it is no longer necessary to use high inertia as energy buffers \cite{gu2022power}. 

Despite the change of circumstances, the emphasis on inertia has been carried over to inverter-based power systems, rather as a tradition than a technical necessity. The requirement of virtual or synthetic inertia has been suggested in grid codes, and the technologies to provide the virtual inertia have been heavily studied in literature \cite{meng2018generalized,fang2018inertia,d2013equivalence,zhong2010synchronverters,li2021revisiting}. On the other hand, there are growing interests in the possibility of low-inertia and even inertia-free power systems. From small-signal analysis, it has been understood that a low inertia-to-droop ratio helps to increase the damping of swing modes in inverter-based systems \cite{pogaku2007modeling,li2022mapping,amin2017small,gu2020impedance,markovic2021understanding}.  
The increased swing damping also extends the transient angle stability region for single-inverter-infinite-bus systems \cite{pan2019transient,fu2020large}. 
Mathematically, it is proved that the synchronisation dynamics of droop-controlled inertia-free grid-forming (GFM) inverters are equivalent to the famous Kuramoto's model \cite{simpson2013synchronization}, which guarantees almost global stability for a multi-inverter system if there exists a proper steady-state power flow \cite{colombino2019global}. This theoretical result is very encouraging but the practical feasibility remains to be established. In particular, it is unknown whether there are excessive transients, e.g. pole slips, in the global trajectory subject to faults. Such excessive transients may exceed the constraints of inverters and grid codes (e.g. over-currents and under-voltages) and threaten system security despite that the trajectory converges eventually.

In this paper, we aim at filling this gap by investigating the first-swing stability of inverter-based inertia-free power systems. 
Compared to global stability, first-swing stability is favoured in practice since it intrinsically prohibits pole slips and avoids the corresponding excessive transients \cite{kundur1994power}. We establish the first-swing stability criterion for inertia-free systems and illustrate that an inertia-free system offers a much wider first-swing stability region than an inertia-rich system. Furthermore, we show that the stability region is fully decentralised and ensures \textit{whole-system stability} for all swing modes across all timescales and all interaction boundaries in a complex grid. This greatly simplifies the transient stability assessment of complex networks and enables simple and unified stability rules to be deployed to distributed resources.

We also discuss the impacts on the frequency stability for inertia-free systems, mainly the frequency ripples and the rate of change of frequency (RoCof) during faults. Such issues can be mitigated by introducing very low but non-zero inertia (i.e. quasi inertia-free) into the system, by changing the frequency droop gain, and by changing RoCof measurement settings in protection devices. The claims of the paper are presented with mathematical proof as well as simulation verification on real-world power systems.


The rest of this paper is organized as follows. The whole-system first-swing stability region of an inertia-free power system is investigated in \sectionref{Section:InertiaFree}. The implications on frequency stability are discussed in \sectionref{Section:Others}. The simulation verification is presented in \sectionref{Section:Simulation}. \sectionref{Section:Conclusion} concludes this paper.


\section{First-Swing Stability of Inertia-Free Systems} \label{Section:InertiaFree}

\begin{figure}
\centering
\includegraphics[scale=1]{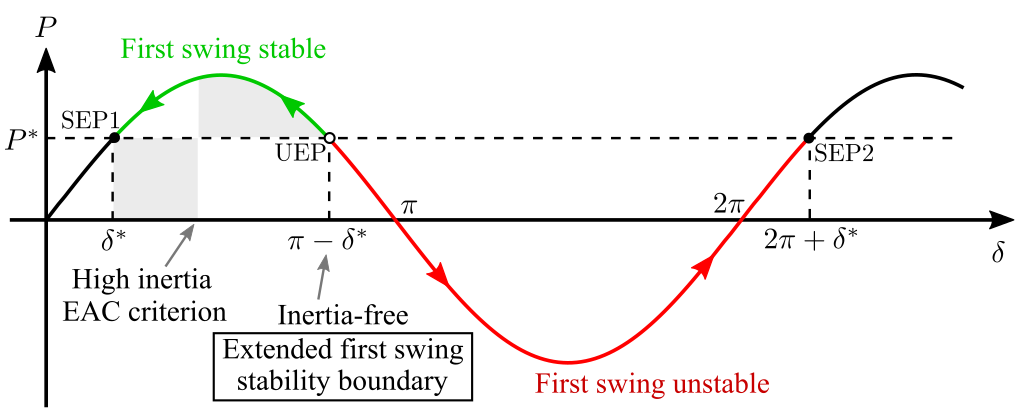}
\caption{First-swing stability boundary for an inertia-free inverter connected to an infinite bus. $\delta$ denotes the angle difference between the inverter and the infinite bus, $\delta^\star$ is the stable equilibrium point (SEP) of $\delta$, $P$ is the electric power injected into the infinite bus, and $P^\star$ is the reference power of the inverter. For an inertia-free system, if the fault is cleared when $\delta<\pi-\delta^*$, the system is pulled back to SEP1 (green trajectory) and is defined as first-swing stable; If the fault is cleared when $\delta>\pi-\delta^*$, the system is pushed away to SEP2 (red trajectory) and is globally transient stable but defined as first-swing unstable.}
\label{Fig:FirstSwingStability}
\end{figure}

Under large grid disturbance, an apparatus is defined as \textit{first-swing stable} if its power angle is always maintained within the unstable equilibrium points (UEPs), and can swing back to its original stable equilibrium point (SEP) [X]. There are two complementary mechanisms can help to regulate system frequency and maintain power balancing. The first is rotating inertia which responds to the derivative of frequency, and the second is primary control (or droop control) which responds to the deviation (error) of frequency. Because of the intrinsic mechanical inertia and slow prime-mover control of a synchronous generator, inertia response dominates the swing dynamics of traditional power systems. Therefore, the equal area criterion (EAC) has been widely used to determine the critical power angle (i.e., first-swing stability boundary) so that enough decelerating area of the rotor can be ensured, as displayed in \figref{Fig:FirstSwingStability}. By contrast, for a grid-forming (GFM) inverter, its frequency regulation is more flexible thanks to its configurable virtual inertia and droop gain \cite{d2013equivalence,li2021revisiting,du2020modeling}. If its primary control is fast enough, its inertia can be replaced. In this case, the first-swing stability is also significantly improved. As illustrated in \figref{Fig:FirstSwingStability}, if inertia is sufficiently small and primary control is sufficiently fast, the first-swing stability region can be pushed beyond the EAC to the boundary of UEP. This is because the primary control has an equivalent damping effect to dissipate the fault-induced kinetic energy almost instantaneously thanks to being inertia-free. This implies that inverters can be endowed with superior rather than inferior stability performances to synchronous generators, by exploiting their controllability.



The intuitive reasoning above can be formalised and generalised to multi-inverter systems by the theorem below. 
\begin{theorem}{}
For a power system with GFM inverters and inductive transmission lines, the region of attraction of the SEP (i.e., the whole-system first-swing stability boundary) extends to the UEP for all inverters if the equivalent inertia of all inverters is sufficiently small.
\end{theorem}

\noindent \textit{Proof}. The state equations for the power system are \cite{kundur1994power,d2013equivalence}
\begin{equation}    \label{eq_sys}
\begin{array}{l}
J_m\dot{\omega}_m=-D_m\omega _m+P_{m}^{\star}-\sum_n{K_{mn}}\sin(\theta _m-\theta _n) \\
\dot \theta_m = \omega_m
\end{array}
\end{equation}
where $\theta$ is angle, $\omega$ is frequency deviation (with respect to the nominal frequency), $J$ is inertia, $D$ is the droop coefficient of the primary control, $P^\star$ is the reference power, $K$ is the synchronization power coefficient determined by the network topology, and the subscripts $m,n$ denote the indices of the $m$th, $n$th inverters. 
The system of (\ref{eq_sys}) can be divided into two coupled sub-systems, one with states $\omega_m$, the other with states $\theta_m$. The $\omega_m$ sub-system has a time constant of $\tau_m = J_m/D_m$. With $J_m$ sufficiently small, $\tau_m$ is also small indicating that $\omega_m$ responds sufficiently fast to the variation of $\theta_m$. This results in a two-timescale system with fast-scale $\omega_m$ and slow-scale $\theta_m$. According to the singular perturbation theory, the fast-scale sub-system can be treated as instantaneous in the slow-scale sub-system \cite{kevorkian2012multiple}. To this end, we let $J_m\rightarrow 0$ \begin{equation}
D_m\omega _m=P_{m}^{\star}-\sum\nolimits_n{K_{mn}}\sin(\theta _m-\theta _n)
\end{equation}
from which follows the reduced system of $\theta_m$
\begin{equation}
\label{eq_reduce}
\dot{\theta}_m=D_{m}^{-1}\,\left( P_{m}^{\star}-\sum\nolimits_n{K_{mn}}\sin(\theta _m-\theta _n) \right).
\end{equation}
Define the Lyapunov function $H$ for the reduced system as
\begin{equation}
H=\sum\nolimits_m{\frac{1}{2}}D_m(\theta _{m}^{\star}-\theta _m)^2
\end{equation}
where $\theta^\star$ is the SEP of $\theta$. The time-derivative of the Lyapunov function is
\begin{equation}
\begin{array}{ll}
\dot{H} \!\!\!\!&=\sum_{m}(\theta_m - \theta_m^\star)\left( P_{m}^{\star}-\sum_n{K_{mn}}\sin(\theta _m-\theta _n) \right)
\\
&=\sum_{m}(\theta_m - \theta_m^\star)\sum_{n} K_{mn}(\sin \delta _{mn}^{\star}-\sin \delta _{mn})
\\
&=\sum_{m<n}{K_{mn}}(\delta _{mn}-\delta _{mn}^{\star})(\sin \delta _{mn}^{\star}-\sin \delta _{mn}) 
\end{array}
\end{equation}
where $\delta_{mn} = \theta_m - \theta_n$, $\delta_{mn}^{\star} = \theta^{\star}_m - \theta^{\star}_n$.


It is clear that
\begin{equation}
\dot{H}<0, \ \forall \ \delta_{mn} \in (-\pi- \delta_{mn}^\star, \pi- \delta_{mn}^\star) \backslash \left\{\delta_{mn}^\star\right\}
\end{equation}
where $\delta_{mn}^\star$ is the SEP and $-\pi- \delta_{mn}^\star$ and $\pi- \delta_{mn}^\star$ are the two closest UEPs. This indicates that $\dot{H}$ is negative-definite within the two UEPs, which proves the theorem. \qed

\begin{figure}
\centering
\includegraphics[scale=1]{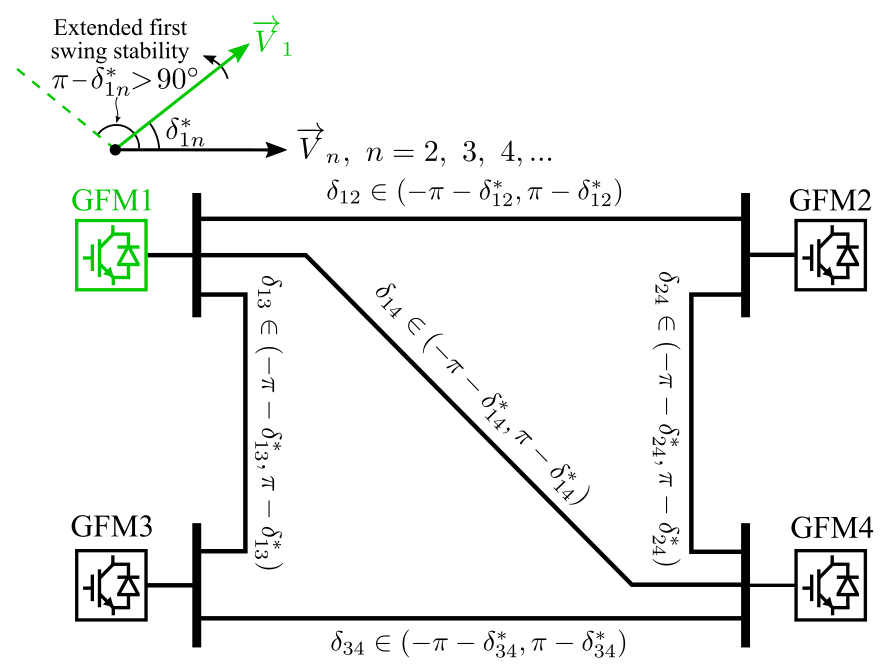}
\caption{First-swing stability boundary of an inverter-based inertia-free power system.}
\label{Fig:FirstSwingStabilityInverterBasedSystem}
\end{figure}

\begin{remark}{} 
\emph{The inertia is sufficiently small if the $\omega_m$ sub-system and $\theta_m$ sub-system are separated in timescale, that is, $\tau_m \ll \lambda^{-1}$ where $\lambda$ is the Lyapunov characteristic exponent of the reduced system (\ref{eq_reduce}) around the transient trajectory ($\lambda$ is an indicator of timescale for a non-linear system) \cite{mease2003timescale}. The time constant $\tau_m$ can be decreased by either decreasing $J_m$ or increasing $D_m$.} 
\end{remark}

\begin{remark}{}
\emph{Theorem 1 has important implications in practice. The SEP $\delta_{mn}^\star$ is bounded within $90^\circ$ due to the property of static power flow, so $(-\pi- \delta_{mn}^\star, \pi- \delta_{mn}^\star) \supset (-\pi/2,\pi/2) $, which means that the transient angle difference can go beyond $90^\circ$ without losing stability, for the m$th$ and n$th$ arbitrary inverter. Moreover, the stability region applies on each pair of angles disregarding the topology of the network. A four-inverter example system is illustrated in \figref{Fig:FirstSwingStabilityInverterBasedSystem}. Thus an inertia-free power system has a significantly extended first-swing stability region and a decentralised stability criterion, i.e., the so-called \textit{whole-system first-swing stability}. This may mitigate the stability barrier of power systems and greatly simplify system operation. It may also introduce a pathway towards new business models of contracting and deploying stability services to distributed resources.}
\end{remark}

\begin{remark}{}
\emph{It is intuitive that an ideally inertia-free system with only positive damping is globally transient stable as long as an SEP exists \cite{simpson2013synchronization}. But it is still worth highlighting that theorem 1 reveals the extended whole-system first-swing stability region and its boundary rather than the global transient stability. As clarified in \figref{Fig:FirstSwingStability}, an inertia-free system can be pulled back to SEP1 (green trajectory, first-swing stable) or pushed away to SEP2  (red trajectory, first-swing unstable), depending on the post-fault point. Even though SEP1 and SEP2 are equivalent at steady state, the system performances are very different. The former one is similar to the conventional re-synchronization procedure. But the latter one suffers reverse power and reverse voltage oscillations. The reverse power (negative power) of an inverter could result in over-voltage risk if the dc-link capacitor is not sufficiently large. The reverse voltage (essentially half-cycle phase shift of the voltage) at apparatus terminal could result in voltage amplitude oscillation along transmission lines and load terminals, as will be illustrated by simulation later. Therefore, it is still useful to find the first-swing stability region, even though the system is globally transient stable.}
\end{remark}

\begin{figure*}[t!]
\centering
\includegraphics[scale=0.72]{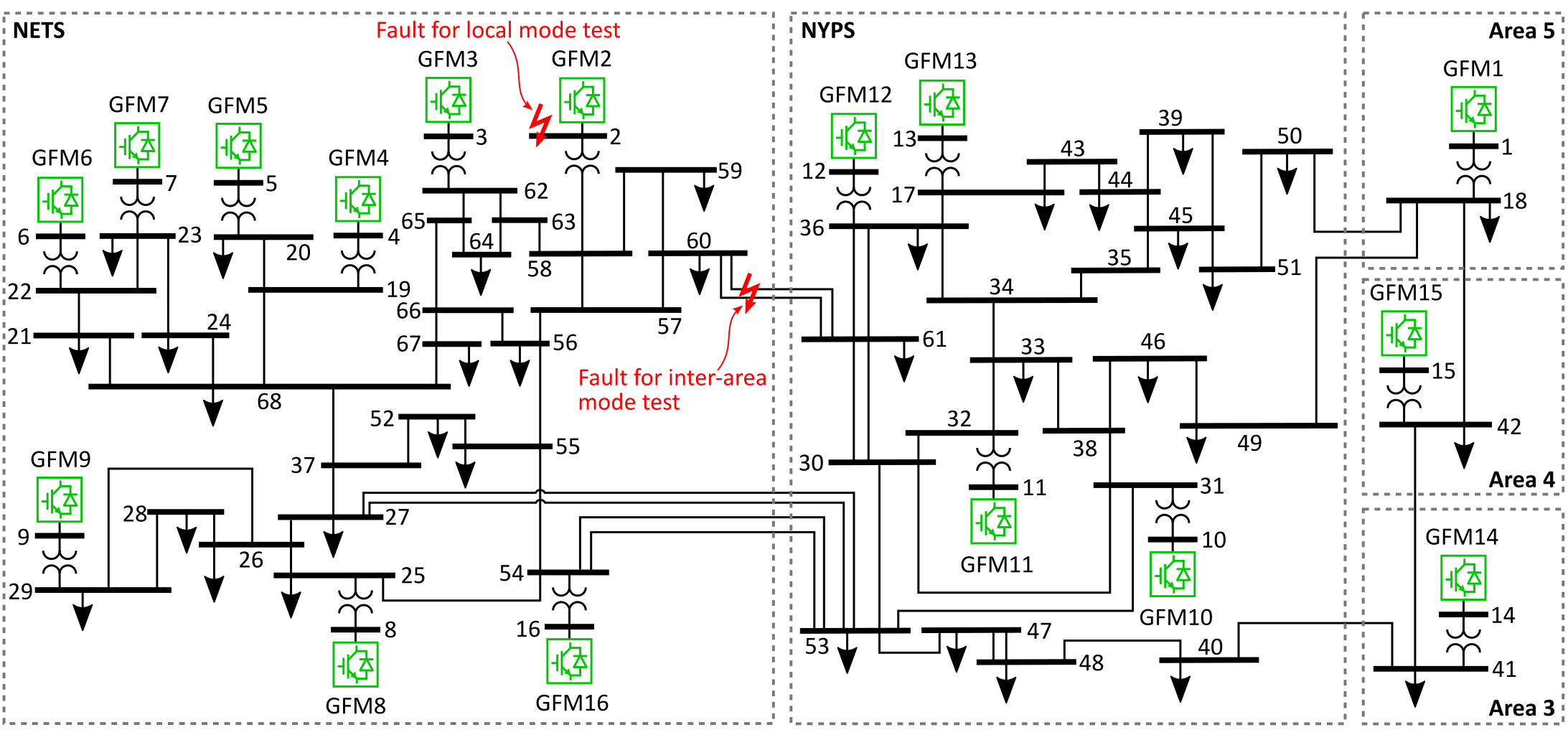}
\caption{Layout of the modified IEEE 68-bus NETS-NYPS power system. All 16 synchronous generators are replaced by by GFM inverters with configurable inertias and droop coefficients.}
\label{Fig:IEEE68Bus}
\end{figure*}




\section{Frequency Stability of Inertia-Free Systems} \label{Section:Others}

Low inertia obviously would lead to large rate-of-change-of-frequency (RoCoF). Synchronous generators are subject to the risk of torque swing and pole slip in the worst case when RoCoF is larger than around 2 Hz/s \cite{uijlings2013independent}, which further leads to apparatus failure. By contrast, a inverter is a full-electrical apparatus and is naturally immune to large RoCoF and frequency deviation. But for an inertia-free inverter, the frequency would be too sensitive to grid disturbance and would suffer large oscillations under grid disturbances, for example, fundamental cycle ripples under short-circuit fault. This could be easily solved by adding a non-zero but still very-low inertia to the inverter \cite{pogaku2007modeling}, i.e., slightly increasing the time constant $\tau=J/D$. This inertia can be physically provided by the dc-link capacitors of inverters with no extra energy storage component needed, thereby is called quasi inertia-free. The effectiveness of this solution will also be validated by simulations in \sectionref{Section:Simulation} later. 

The conventional protection system based on RoCoF may also need to be updated in an inertia-free 100\%-inverter system \cite{uijlings2013independent,he2021transient}. A promising alternative to replace frequency derivative protection is frequency deviation protection, but which may rely on an accurate measurement of frequency and needs to be further investigated in the future \cite{anderson2022power}.

In addition to GFM inverters, inverter-based power systems may also have grid-following (GFL) inverters. The GFL inverters can be approximately regarded as constant power sources or loads within the phase-locked loop (PLL) bandwidth, which can also be directly compensated by the fast primary control of GFM inverters during the frequency regulation. As will be validated in \sectionref{Section:Simulation} later, it is still beneficial to pursue inertia-free in a hybrid GFM-GFL power system. It should also be noted that, even in a zero-carbon grid, the existence of legacy synchronous generators for hydro and nuclear plants would provide still intrinsic inertias. The hybrid connection of inertia and inertia-free systems also needs to be further investigated in future work \cite{he2021transient}. 



\section{Simulation Results}    \label{Section:Simulation}

\begin{figure*}[t!]
\centering
\includegraphics[scale=0.72]{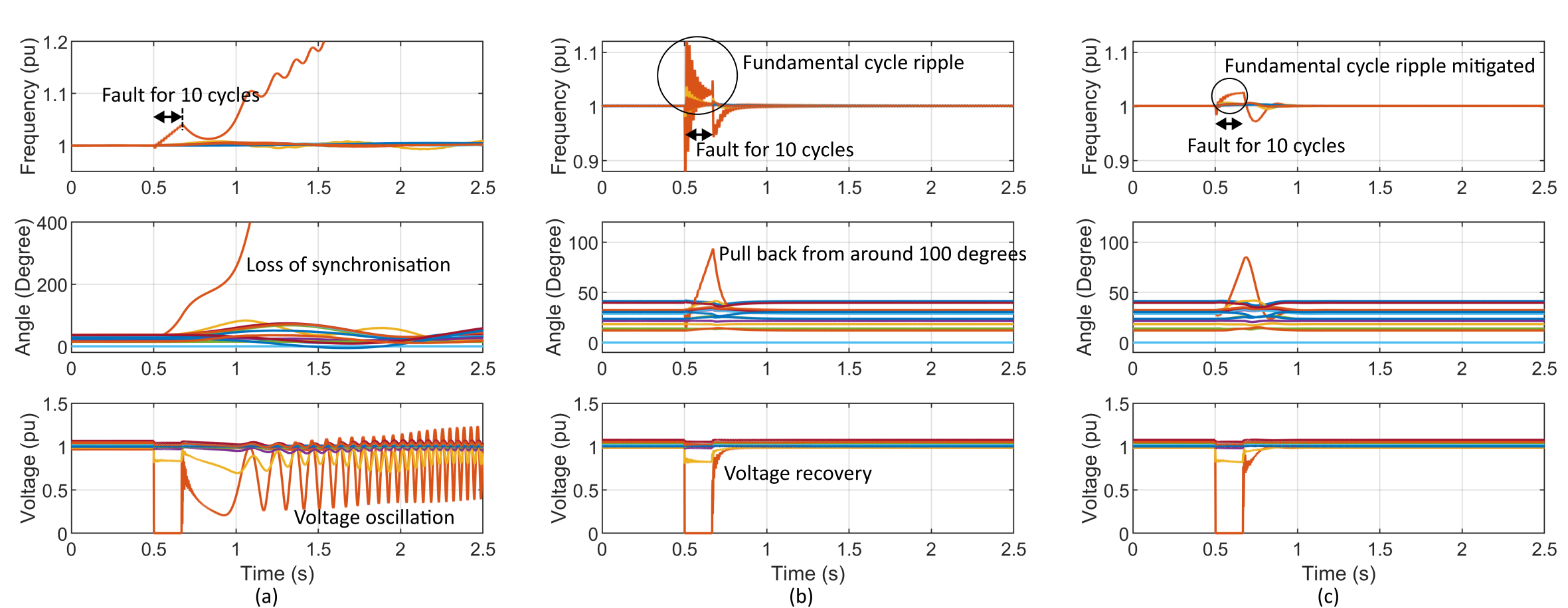}
\caption{Local mode transient responses of the full-GFM system subject to a symmetric (three-phase-to-ground) short-circuit fault near bus 2. (a) High inertia and low droop coefficient, $\tau = 80~\text{s}$. (b) Inertia-free, $\tau = 1.33~\text{ms}$. (c) Quasi inertia-free, $\tau = 40~\text{ms}$.}
\label{Fig:LocalMode}
\end{figure*}

\begin{figure*}[t!]
\centering
\includegraphics[scale=0.72]{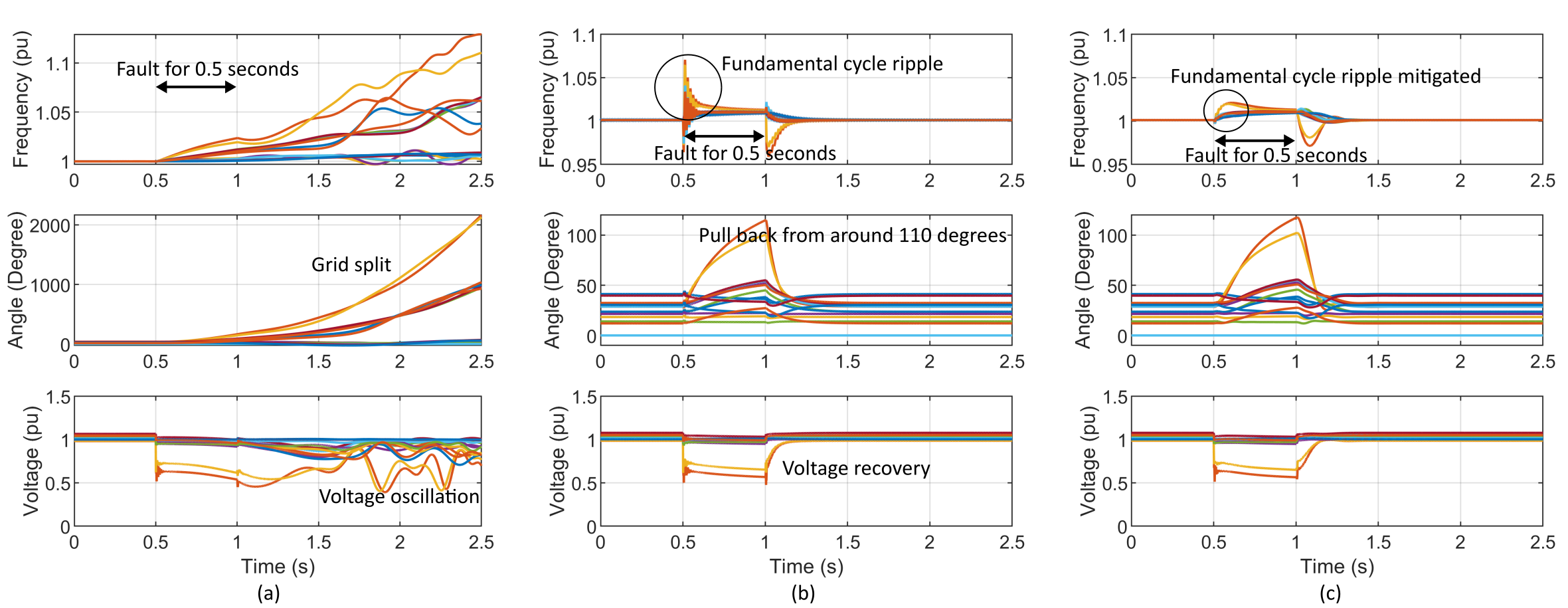}
\caption{Inter-area mode transient responses of the full-GFM system subject to a symmetric (three-phase-to-ground) short-circuit fault near bus 60. (a) High inertia and low droop coefficient, $\tau = 80~\text{s}$. (b) Inertia-free, $\tau = 1.33~\text{ms}$. (c) Quasi inertia-free, $\tau = 40~\text{ms}$.}
\label{Fig:InterAreaMode}
\end{figure*}

The IEEE 68-bus NETS-NYPS power systems in \figref{Fig:IEEE68Bus} (full-GFM) and \figref{Fig:IEEE68BusHybrid} (hybrid-GFM-GFL) are tested next. Matlab/Simulink is used for electromagnetic transient (EMT) simulation of the test systems. The parameters and simulation models are available online \cite{FuturePowerNetworks}. 
Either symmetric or asymmetric short-circuit fault is applied near either bus 2 or bus 60 to trigger local and inter-area angle swing. Excessive fault-clearing times (10 cycles for local mode and 0.5 s for inter-area mode) are used to test the stability limit. All GFM inverters are configured to have a uniform time constant $\tau = J/D$ varying from 80 s to 1.33 ms in different cases. All GFL inverters are configured with a fixed PLL bandwidth of 10 Hz.

\subsection{Full-GFM System: Symmetric Fault Test}

The symmetric (three-phase-to-ground) fault test results in \figref{Fig:LocalMode} and \figref{Fig:InterAreaMode} are discussed next first. Three cases (a)-(c) are tested with different combinations of inertia $J$ and droop coefficient $D$. The case (a) with high inertia and low droop coefficient ($\tau = 80~\text{s}$) experiences loss of synchronization for GFM2 near the fault in the local mode (\figref{Fig:LocalMode}), or grid-splitting along the faulted interconnector in the inter-area mode (\figref{Fig:InterAreaMode}). In contrast, the system remains stable subject to the same faults if inertias are reduced and droop coefficients are increased, as illustrated in cases (b)-(c). The angles are pulled back from beyond $90^\circ$ and re-synchronized after faults for both the local and inter-area swing modes. This verifies the superior performance of angle stability for the inertia-free power system. The stability is consistent for local and inter-area modes and thus scalable to complex networks. 

\begin{figure}[t!]
\centering
\includegraphics[scale=0.5]{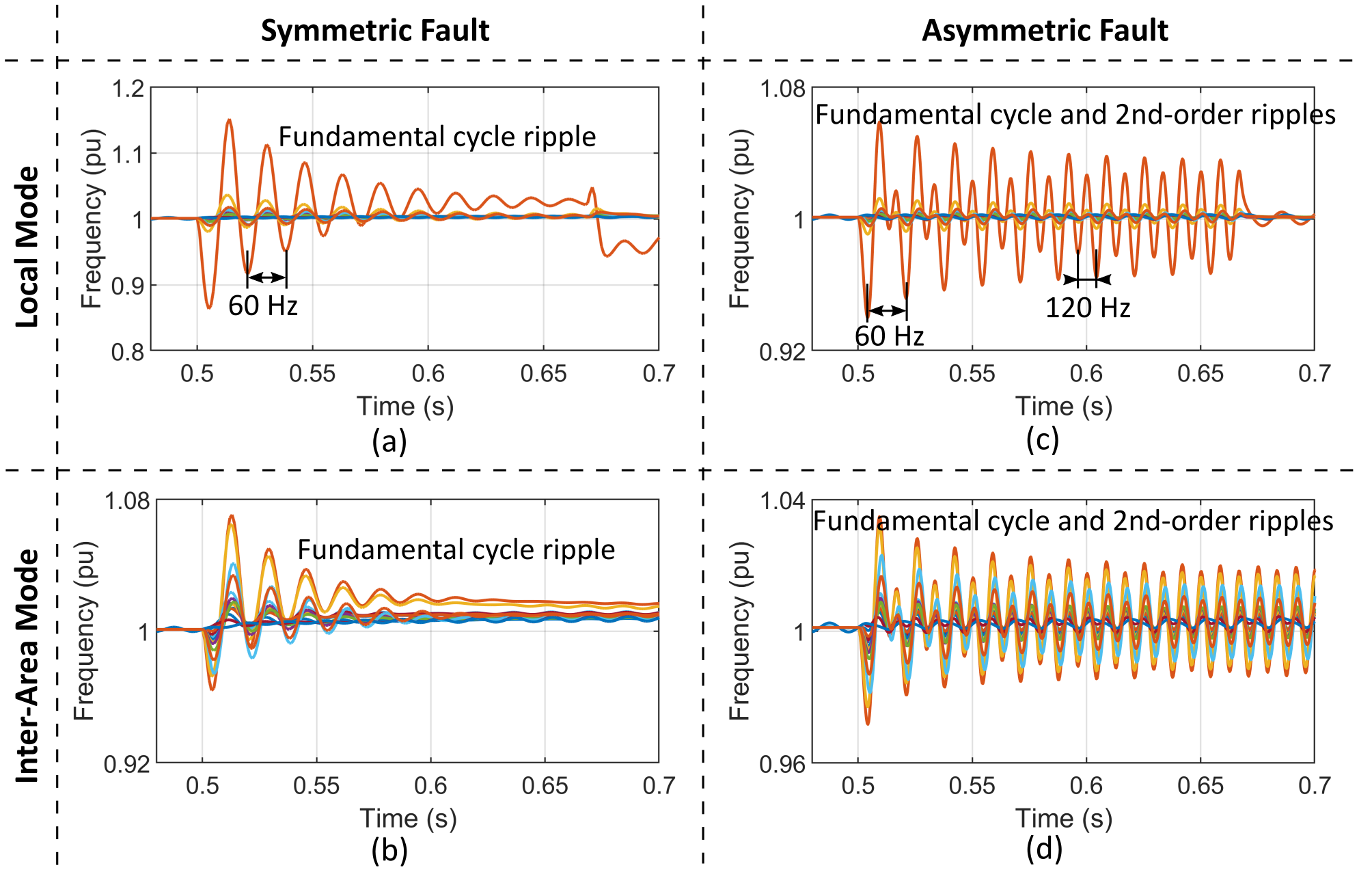}
\caption{Zoomed-in waveform (0.5 s to 0.7 s) of frequency for inertia-free system under different fault types. (a) Symmetric fault, local mode, in \figref{Fig:LocalMode}b. (b) Symmetric fault, inter-area mode, in \figref{Fig:InterAreaMode}b. (c) Asymmetric fault, local mode, in \figref{Fig:LocalModeUnbFault}b. (d) Asymmetric fault, inter-area mode, in \figref{Fig:InterAreaModeUnbFault}b.}
\label{Fig:ZoomIn}
\end{figure}

A side effect is observed for case (b) (almost zero inertia, $\tau = 1.33~\text{ms}$) in both \figref{Fig:LocalMode} and \figref{Fig:InterAreaMode}, which is the fundamental cycle ripple in the frequency during the fault transients \cite{kundur1994power}, as zoomed in \figref{Fig:ZoomIn}(a) and (b). This is induced by that the power ripple during fault transient is fed through to frequency directly when inertia is low. We can mitigate this ripple by slightly increasing the inertia, as is illustrated in the quasi inertia-free case (c). In this case, the time constant is set to $\tau = 40~\text{ms}$ which is still sufficiently small compared to the timescale of angle swing (on the scale of a second seen from \figref{Fig:LocalMode} and \figref{Fig:InterAreaMode}), but is large enough to filter the fundamental cycle ripple. 

Large voltage oscillations can be continuously observed after the fault is cleared in case (a) for both \figref{Fig:LocalMode} and \figref{Fig:InterAreaMode}. As discussion in Remark 3 in \sectionref{Section:InertiaFree}, this phenomenon happens because of the loss of synchronization of apparatuses and the continuously-rotating voltage vector beyond first-swing boundary at these apparatus terminals. This further implies the importance of first-swing stability and the benefit of inertia-free.

\subsection{Full-GFM System: Asymmetric Fault Test}

\begin{figure*}[t!]
\centering
\includegraphics[scale=0.72]{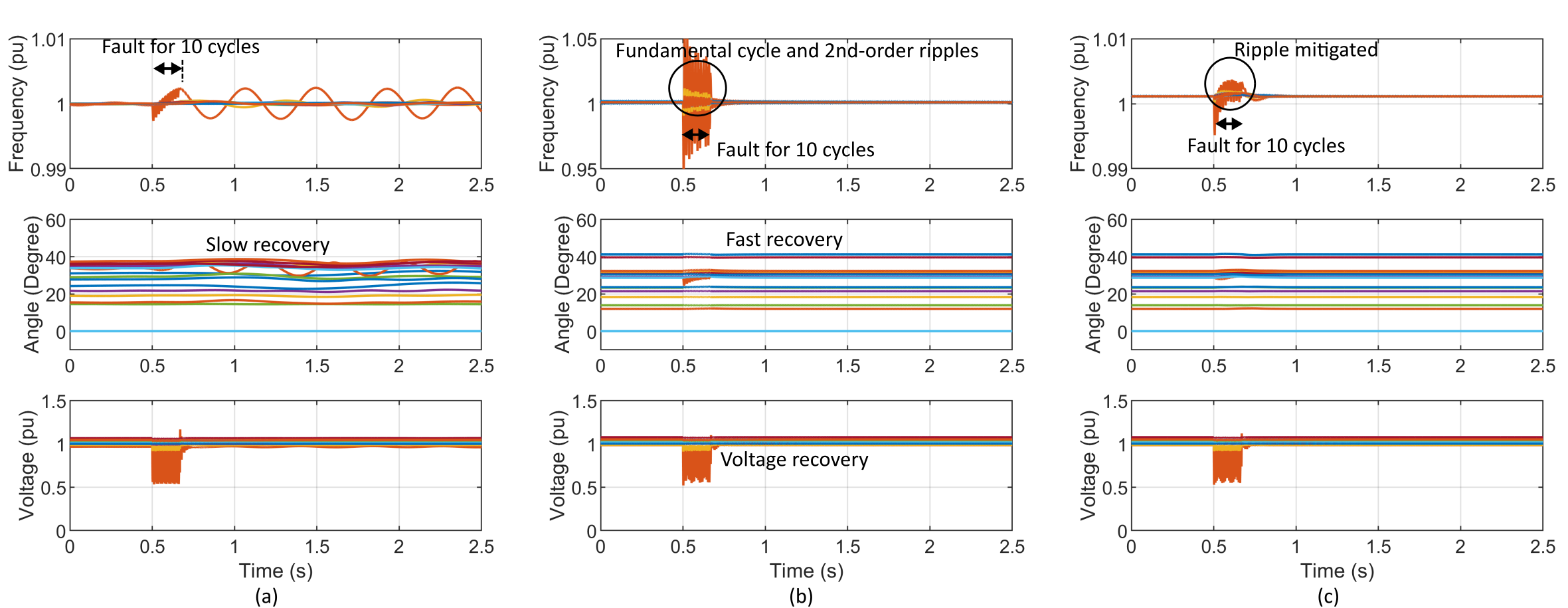}
\caption{Local mode transient responses of the full-GFM system subject to an asymmetric (single-phase-to-ground) short-circuit fault  near bus 2. (a) High inertia and low droop coefficient, $\tau = 80~\text{s}$. (b) Inertia-free, $\tau = 1.33~\text{ms}$. (c) Quasi inertia-free, $\tau = 40~\text{ms}$.}
\label{Fig:LocalModeUnbFault}
\end{figure*}

\begin{figure*}[t!]
\centering
\includegraphics[scale=0.72]{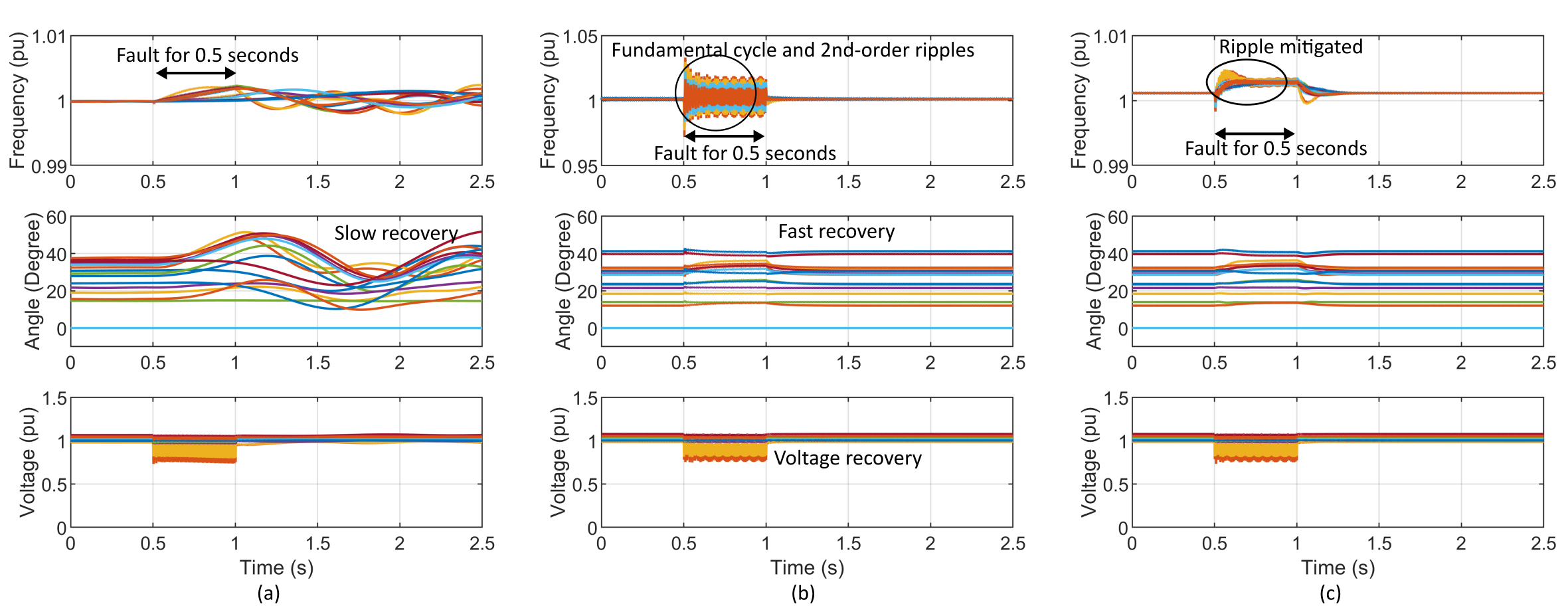}
\caption{Inter-area mode transient responses of the full-GFM system subject to an asymmetric (single-phase-to-ground) short-circuit fault near bus 60. (a) High inertia and low droop coefficient, $\tau = 80~\text{s}$. (b) Inertia-free, $\tau = 1.33~\text{ms}$. (c) Quasi inertia-free, $\tau = 40~\text{ms}$.}
\label{Fig:InterAreaModeUnbFault}
\end{figure*}

\begin{figure*}[t!]
\centering
\includegraphics[scale=0.72]{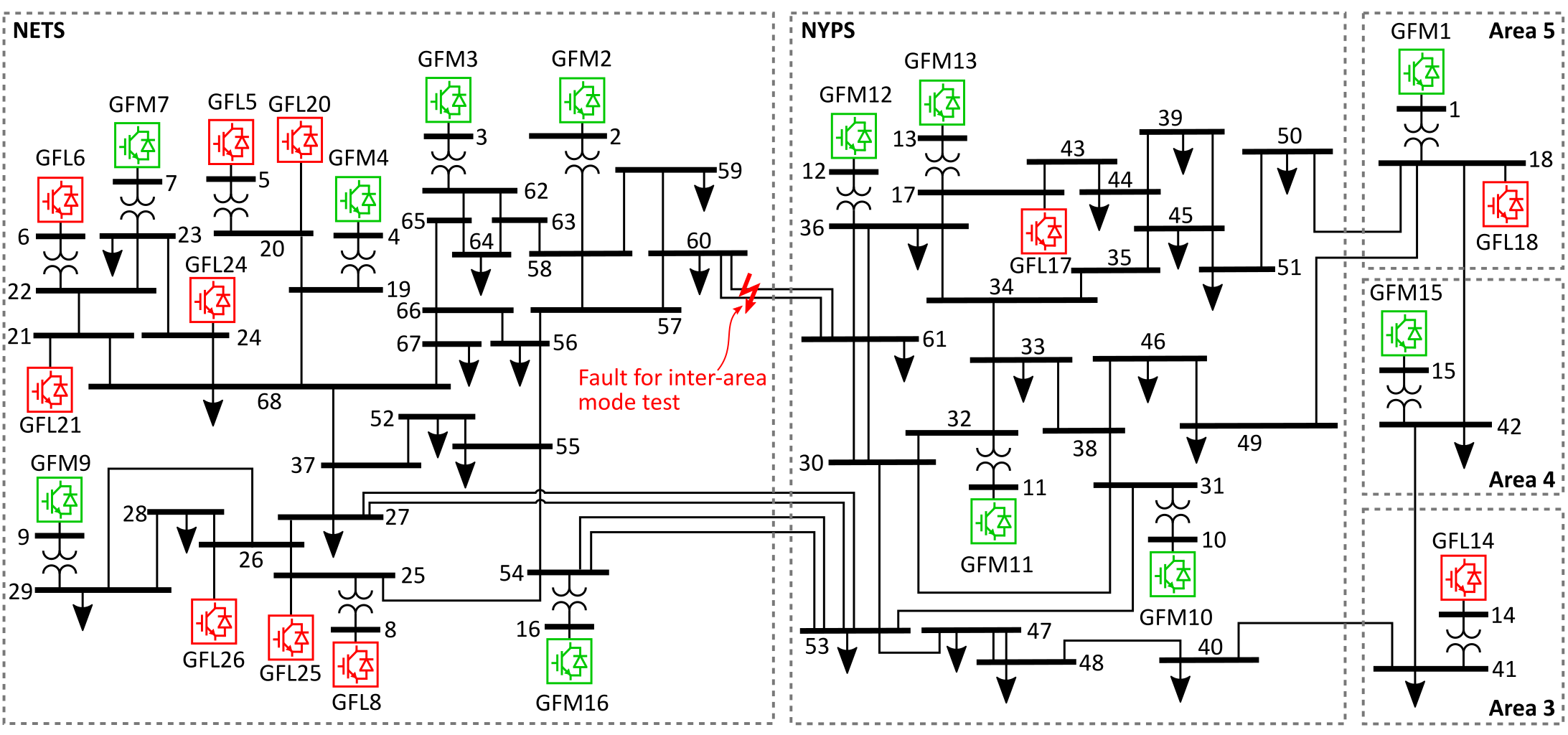}
\caption{Layout of the modified IEEE 68-bus NETS-NYPS power system. The hybrid system consists of 12 GFM inverters (green) and 12 GFL inverters (red).}
\label{Fig:IEEE68BusHybrid}
\end{figure*}

\begin{figure*}[t!]
\centering
\includegraphics[scale=0.72]{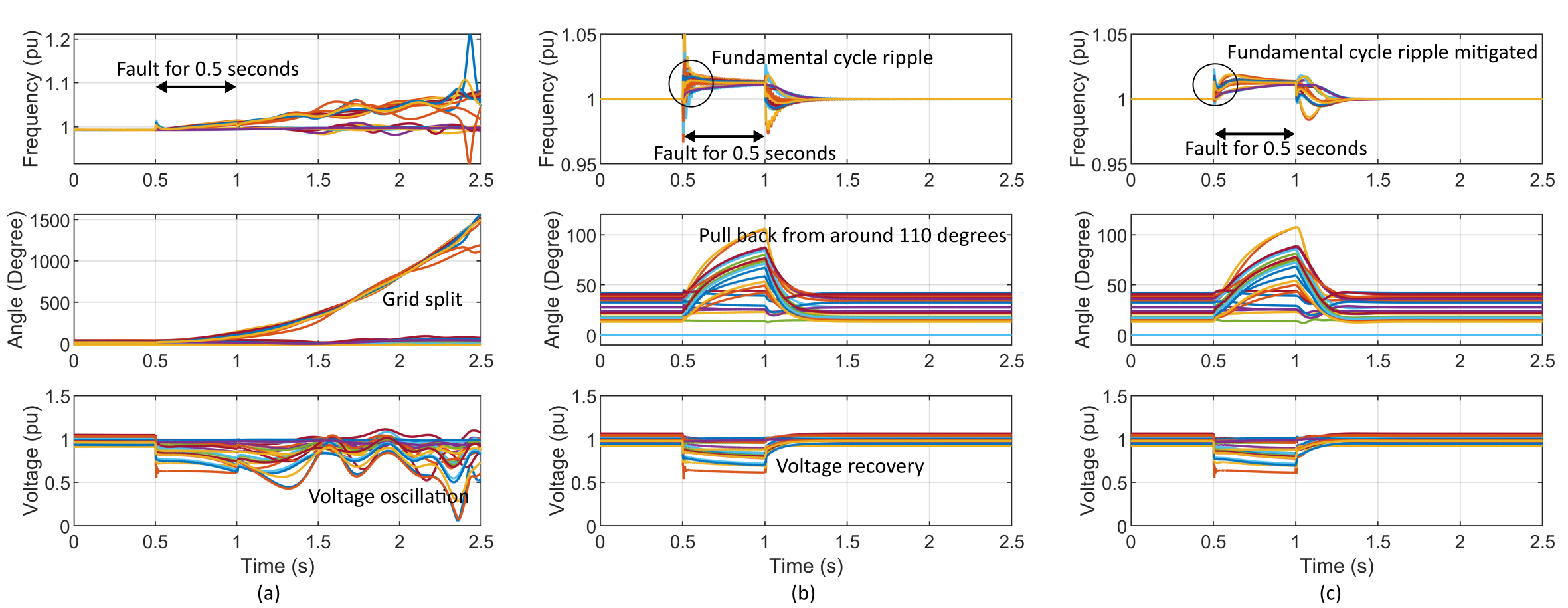}
\caption{Inter-area mode transient responses of the hybrid GFM-GFL system subject to an symmetric (three-phase-to-ground) short-circuit fault near bus 60. (a) High inertia and low droop coefficient, $\tau = 80~\text{s}$. (b) Inertia-free, $\tau = 1.33~\text{ms}$. (c) Quasi inertia-free, $\tau = 40~\text{ms}$.}
\label{Fig:InterAreaModeHybrid}
\end{figure*}

The asymmetric (single-phase-to-ground) fault test results in \figref{Fig:LocalModeUnbFault} and \figref{Fig:InterAreaModeUnbFault} are discussed next. The asymmetric fault is less severe than the symmetric fault, therefore all cases (a), (b), (c) here do not loss the synchronization after the fault is cleared. But, the high-inertia and low-droop-coefficient case in (a) ($\tau = 80~\text{s}$) suffers almost equal-amplitude oscillation after the fault is cleared because of the very large $\tau=J/D$. By contrast, the inertia-free case in (b) ($\tau = 1.33~\text{ms}$) recovers much faster after the fault is cleared, which implies the enhanced transient stability. The during-fault frequency oscillations occur in the inertia-free case (b) again. But differing from the symmetric fault case, the ripple here is jointly dominated by fundamental cycle (60 Hz) and its 2nd-order (120 Hz) ripples, as compared in \figref{Fig:Zoomin}. The 2nd-order ripple is fed by the power oscillations under the asymmetric fault event. Both the fundamental and 2nd-order ripples here can be mitigated by adding a very-low but non-zero inertia, as illustrated in the quasi inertia-free case in (c) ($\tau = 40~\text{ms}$).

\subsection{Hybrid GFM-GFL System}

Next, to test the hybrid GFM-GFL system, we replace 4 GFM inverters (at buses 5, 6, 8 and 14) by GFL inverters, and replace 8 passive loads (at buses 17, 18, 20, 21, 23, 24, 25, 26) also by GFL-type active loads. This gives a modified IEEE 68-bus power system with 12 GFM inverters and 12 GFL inverters, as shown in the layout in \figref{Fig:IEEE68BusHybrid}. The test results of the new system with different inertia configurations of GFMs are shown in \figref{Fig:InterAreaModeHybrid}. The high-inertia case in (a) ($\tau = 80~\text{s}$) experiences the loss of synchronization and the grid splitting. By contrast, the inertia-free case in (b) ($\tau = 1.33~\text{ms}$) remains stable after the fault is cleared. The fundamental cycle ripple in (b) can be mitigated by slightly increasing the inertia in the quasi inertia-free case in (c) ($\tau = 40~\text{ms}$). It should be noted that the results of hybrid GFM-GFL system in \figref{Fig:InterAreaModeHybrid} here are very similar to the results of full-GFM system in \figref{Fig:InterAreaMode} in previous subsection, which validates the feasibility of theory also in hybrid GFM-GFL system.


\section{Conclusions}   \label{Section:Conclusion}
An inertia-free power system driven by inverter-based resources is both feasible and beneficial. Replacing inertia by fast primary control as the major mechanism of frequency regulation extends the first-swing stability region beyond $90^\circ$ for all inverters and decentralises the stability criterion, i.e., the so-called whole-system first-swing stability. As verified by simulations of IEEE 68-bus power system, a inertia-free system is more robust than a high-inertia system under both symmetric and asymmetric short-circuit faults.

\ifCLASSOPTIONcaptionsoff
  \newpage
\fi

\bibliographystyle{IEEEtran}
\bibliography{Paper}

\begin{thebibliography}{10}
\providecommand{\url}[1]{#1}
\csname url@samestyle\endcsname
\providecommand{\newblock}{\relax}
\providecommand{\bibinfo}[2]{#2}
\providecommand{\BIBentrySTDinterwordspacing}{\spaceskip=0pt\relax}
\providecommand{\BIBentryALTinterwordstretchfactor}{4}
\providecommand{\BIBentryALTinterwordspacing}{\spaceskip=\fontdimen2\font plus
\BIBentryALTinterwordstretchfactor\fontdimen3\font minus
  \fontdimen4\font\relax}
\providecommand{\BIBforeignlanguage}[2]{{%
\expandafter\ifx\csname l@#1\endcsname\relax
\typeout{** WARNING: IEEEtran.bst: No hyphenation pattern has been}%
\typeout{** loaded for the language `#1'. Using the pattern for}%
\typeout{** the default language instead.}%
\else
\language=\csname l@#1\endcsname
\fi
#2}}
\providecommand{\BIBdecl}{\relax}
\BIBdecl

\bibitem{kundur1994power}
P.~Kundur, N.~J. Balu, and M.~G. Lauby, \emph{{Power system stability and
  control}}.\hskip 1em plus 0.5em minus 0.4em\relax McGraw-hill New York, 1994,
  vol.~7.

\bibitem{hatziargyriou2020definition}
N.~Hatziargyriou, J.~Milanovic, C.~Rahmann, V.~Ajjarapu, C.~Canizares,
  I.~Erlich, D.~Hill, I.~Hiskens, I.~Kamwa, B.~Pal \emph{et~al.}, ``{Definition
  and classification of power system stability revisited \& extended},''
  \emph{IEEE Transactions on Power Systems}, 2020.

\bibitem{markovic2021understanding}
U.~Markovic, O.~Stanojev, P.~Aristidou, E.~Vrettos, D.~Callaway, and G.~Hug,
  ``Understanding small-signal stability of low-inertia systems,'' \emph{IEEE
  Transactions on Power Systems}, vol.~36, no.~5, pp. 3997--4017, 2021.

\bibitem{gu2022power}
Y.~Gu and T.~C. Green, ``Power system stability with a high penetration of
  inverter-based resources,'' \emph{Proceedings of the IEEE}, 2022.

\bibitem{rocabert2012control}
J.~Rocabert, A.~Luna, F.~Blaabjerg, and P.~Rodriguez, ``{Control of power
  converters in AC microgrids},'' \emph{IEEE transactions on power
  electronics}, vol.~27, no.~11, pp. 4734--4749, 2012.

\bibitem{o2021enabling}
M.~O’Malley, T.~Bowen, J.~Bialek, M.~Braun, N.~Cutululis, T.~Green,
  A.~Hansen, E.~Kennedy, J.~Kiviluoma, J.~Leslie \emph{et~al.}, ``Enabling
  power system transformation globally: a system operator research agenda for
  bulk power system issues,'' \emph{IEEE Power and Energy Magazine}, vol.~19,
  no.~6, pp. 45--55, 2021.

\bibitem{fang2018inertia}
J.~Fang, H.~Li, Y.~Tang, and F.~Blaabjerg, ``On the inertia of future
  more-electronics power systems,'' \emph{IEEE Journal of Emerging and Selected
  Topics in Power Electronics}, vol.~7, no.~4, pp. 2130--2146, 2018.

\bibitem{li2021revisiting}
Y.~Li, Y.~Gu, and T.~C. Green, ``{Revisiting Grid-Forming and Grid-Following
  Inverters: A Duality Theory},'' \emph{IEEE Transactions on Power Systems},
  2022.

\bibitem{meng2018generalized}
X.~Meng, J.~Liu, and Z.~Liu, ``A generalized droop control for grid-supporting
  inverter based on comparison between traditional droop control and virtual
  synchronous generator control,'' \emph{IEEE Transactions on Power
  Electronics}, vol.~34, no.~6, pp. 5416--5438, 2018.

\bibitem{d2013equivalence}
S.~D'Arco and J.~A. Suul, ``{Equivalence of virtual synchronous machines and
  frequency-droops for converter-based microgrids},'' \emph{IEEE Transactions
  on Smart Grid}, vol.~5, no.~1, pp. 394--395, 2013.

\bibitem{zhong2010synchronverters}
Q.-C. Zhong and G.~Weiss, ``Synchronverters: Inverters that mimic synchronous
  generators,'' \emph{IEEE transactions on industrial electronics}, vol.~58,
  no.~4, pp. 1259--1267, 2010.

\bibitem{pogaku2007modeling}
N.~Pogaku, M.~Prodanovic, and T.~C. Green, ``{Modeling, analysis and testing of
  autonomous operation of an inverter-based microgrid},'' \emph{IEEE
  Transactions on power electronics}, vol.~22, no.~2, pp. 613--625, 2007.

\bibitem{li2022mapping}
Y.~Li, Y.~Gu, and T.~C. Green, ``Mapping of dynamics between mechanical and
  electrical ports in sg-ibr composite grids,'' \emph{IEEE Transactions on
  Power Systems}, vol.~37, no.~5, pp. 3423--3433, 2022.

\bibitem{amin2017small}
M.~Amin and M.~Molinas, ``{Small-signal stability assessment of power
  electronics based power systems: A discussion of impedance-and
  eigenvalue-based methods},'' \emph{IEEE Transactions on Industry
  Applications}, vol.~53, no.~5, pp. 5014--5030, 2017.

\bibitem{gu2020impedance}
Y.~Gu, Y.~Li, Y.~Zhu, and T.~C. Green, ``{Impedance-based whole-system modeling
  for a composite grid via embedding of frame dynamics},'' \emph{IEEE
  Transactions on Power Systems}, vol.~36, no.~1, pp. 336--345, 2020.

\bibitem{pan2019transient}
D.~Pan, X.~Wang, F.~Liu, and R.~Shi, ``Transient stability of voltage-source
  converters with grid-forming control: A design-oriented study,'' \emph{IEEE
  Journal of Emerging and Selected Topics in Power Electronics}, vol.~8, no.~2,
  pp. 1019--1033, 2019.

\bibitem{fu2020large}
X.~Fu, J.~Sun, M.~Huang, Z.~Tian, H.~Yan, H.~H.-C. Iu, P.~Hu, and X.~Zha,
  ``Large-signal stability of grid-forming and grid-following controls in
  voltage source converter: A comparative study,'' \emph{IEEE Transactions on
  Power Electronics}, vol.~36, no.~7, pp. 7832--7840, 2020.

\bibitem{simpson2013synchronization}
J.~W. Simpson-Porco, F.~D{\"o}rfler, and F.~Bullo, ``Synchronization and power
  sharing for droop-controlled inverters in islanded microgrids,''
  \emph{Automatica}, vol.~49, no.~9, pp. 2603--2611, 2013.

\bibitem{colombino2019global}
M.~Colombino, D.~Gro{\ss}, J.-S. Brouillon, and F.~D{\"o}rfler, ``Global phase
  and magnitude synchronization of coupled oscillators with application to the
  control of grid-forming power inverters,'' \emph{IEEE Transactions on
  Automatic Control}, vol.~64, no.~11, pp. 4496--4511, 2019.

\bibitem{du2020modeling}
W.~Du, F.~K. Tuffner, K.~P. Schneider, R.~H. Lasseter, J.~Xie, Z.~Chen, and
  B.~Bhattarai, ``{Modeling of grid-forming and grid-following inverters for
  dynamic simulation of large-scale distribution systems},'' \emph{IEEE
  Transactions on Power Delivery}, vol.~36, no.~4, pp. 2035--2045, 2020.

\bibitem{kevorkian2012multiple}
J.~K. Kevorkian and J.~D. Cole, \emph{Multiple scale and singular perturbation
  methods}.\hskip 1em plus 0.5em minus 0.4em\relax Springer Science \& Business
  Media, 2012, vol. 114.

\bibitem{mease2003timescale}
K.~Mease, S.~Bharadwaj, and S.~Iravanchy, ``Timescale analysis for nonlinear
  dynamical systems,'' \emph{Journal of guidance, control, and dynamics},
  vol.~26, no.~2, pp. 318--330, 2003.

\bibitem{uijlings2013independent}
W.~Uijlings, D.~Street, and S.~London, ``An independent analysis on the ability
  of generators to ride through rate of change of frequency values up to
  2hz/s,'' \emph{EirGrid, London, UK, Rep}, vol. 16010927, 2013.

\bibitem{he2021transient}
X.~He, S.~Pan, and H.~Geng, ``Transient stability of hybrid power systems
  dominated by different types of grid-forming devices,'' \emph{IEEE
  Transactions on Energy Conversion}, vol.~37, no.~2, pp. 868--879, 2021.

\bibitem{anderson2022power}
P.~M. Anderson, C.~F. Henville, R.~Rifaat, B.~Johnson, and S.~Meliopoulos,
  \emph{Power system protection}.\hskip 1em plus 0.5em minus 0.4em\relax John
  Wiley \& Sons, 2022.

\bibitem{FuturePowerNetworks}
\BIBentryALTinterwordspacing
``{Future Power Networks}.'' [Online]. Available:
  \url{https://github.com/Future-Power-Networks/Publications}
\BIBentrySTDinterwordspacing

\end{thebibliography}

\end{document}